\documentclass[a4paper]{article}
\usepackage{dx-2013}
\usepackage[T1]{fontenc}
\usepackage{ae,aecompl}
\usepackage{amsfonts}
\usepackage{amsmath}
\usepackage{amsthm}

\usepackage{times}

\usepackage{graphicx}      
\usepackage{amssymb}
\usepackage{subfigure}
\usepackage{tikz}

\usepackage{savesym}
\savesymbol{AND}
\savesymbol{OR}
\savesymbol{NOT}
\savesymbol{TO}
\savesymbol{COMMENT}
\savesymbol{BODY}
\savesymbol{IF}
\savesymbol{ELSE}
\savesymbol{ELSIF}
\savesymbol{FOR}
\savesymbol{WHILE}
\usepackage{algorithmic}
\usepackage{algorithm}
\usepackage{xspace}

\usetikzlibrary{automata,petri,positioning}
\usetikzlibrary{shapes,snakes,arrows, automata,shadows,calc}
\tikzstyle{state}=[circle,draw,thick,inner sep=1.5mm]

\newcommand{\diag}[1]{\textbf{diag}(#1)}

\newcommand{\imp}{\Rightarrow}

\newcommand{\trace}[1]{\textrm{trace}(#1)}
\newcommand{\camino}[1]{\textrm{trace}^{-1}(#1)}

\newcommand{\obs}[1]{obs(#1)}

\newcommand{\proj}[2]{P_{#1}({#2})}

\newcommand{\buchi}                {\mbox{B\"uchi}\xspace}
\newcommand{\eventually} {\Diamond\,}
\newcommand{\A}[1][\neg\varphi]    {\mbox{$A_{#1}$}\xspace}
\newcommand{\N}[1][\neg\varphi]    {\mbox{$\netn_{#1}$}\xspace}
\newcommand{\veri}  {\ensuremath{\mathcal{V}}\xspace}

\newcommand{\ndiagltl}   {\ensuremath{\overline{\bf diag}\xspace}}

\newcommand{\DEF}            {\stackrel{\mbox{{\tiny\rm df}}}{=}}
\newcommand{\punf}           {\textsc{Punf}\xspace}

\renewcommand\alph{\mathbb{A}}

\newcommand\by{\begin{eqnarray}}
\newcommand\ey{\end{eqnarray}}

\newcommand\bys{\begin{eqnarray*}}
\newcommand\eys{\end{eqnarray*}}

\newcommand\bdf{\begin{definition}}
\newcommand\edf{\end{definition}}
\newcommand\bth{\begin{theorem}}
\newcommand\ethm{\end{theorem}}
\newcommand\bel{\begin{lemma}}
\newcommand\enl{\end{lemma}}
\newcommand\bum{\begin{enumerate}}
\newcommand\eum{\end{enumerate}}
\newcommand\bit{\begin{itemize}}
\newcommand\eit{\end{itemize}}

\newcommand\move[1]{\stackrel{#1}{\longrightarrow }}
\newcommand\onlynetn{{\mathit{N}}}
\newcommand{\places}{\mathit{P}}

\newcommand\placen{\mathit{p}}
\newcommand\transn{\mathit{t}}
\newcommand\trans{{\mathit{T}}}
\newcommand\netn{{\mathcal{N}}}

\newcommand\flow{\mathit{F}}

\newcommand{\preset}[1]{{{}^\bullet{#1}}}

\newcommand{\postset}[1]{{#1}^\bullet}
\newcommand{\markn}{\mathit{M}}

\newcommand\reach{\mathbf{R}}

\newtheorem{assum}{Assumption}
\newtheorem{exmp}{Example}
\newtheorem{prop}{Proposition}
\newtheorem{defn}{Definition}
\newtheorem{thm}{Theorem}
\newtheorem{pf}{Proof}

\newcommand{\runs}[1]{\mathrm{run}(#1)}

\newcommand{\inftraces}[1]{\mathrm{traces}^{\omega}(#1)}

\title{Distributed Diagnosability Analysis with Petri Nets
\thanks{This work has been supported by the European Union Sh Framework Program under grant agreement no. 295261 (MEALS), and
the project SticAmSud No. 13STIC-04.}}
\author%
{%
L. Brand\'an-Briones$^1$ \and A. Madalinski$^2$ \and H. Ponce-de-Le\'on$^3$\\
$^1$CONICET and Fa.M.A.F. - Universidad Nacional de C\'ordoba, Argentina\\
$^2$Facultad de Ciencias de la Ingenier\'ia, Universidad Austral de Chile, Valdivia, Chile\\
$^2$INRIA and LSV, \'Ecole Normale Sup\'erieure de Cachan and CNRS, France\\
}

\begin{document}

\maketitle

\begin{abstract}                
We propose a framework to distributed diagnosability analysis of concurrent systems modeled with Petri nets as a collection of components synchronizing on common observable transitions, where faults can occur in several components. The diagnosability analysis of the entire system is done in parallel by verifying the interaction of each component with the fault free versions of the other components. Furthermore, we use existing efficient methods and tools, in particular parallel LTL-X model checking based on unfoldings, for diagnosability verification.


\end{abstract}


\section{Introduction}

As systems become larger their behavior becomes more and more complex, consequently it becomes harder to detect faults. There are cases where faults cannot be ruled out at design stage since they intrinsically belong to the systems (they are inherent faults), or to the environment where the system is executed. Therefore, it becomes crucially important to have mechanisms in place to be able to detect and recover from such faults when they occur.

In the last years a lot of work have been done studying inherent faults: \emph{Fault diagnosis} consists in detecting abnormal behaviors of a physical system. \emph{Diagnosability} is the property that gives the possibility of detecting faults in a bounded time after they occur given a set of observations. If a system is diagnosable, it is always possible to determine if a fault has occurred by observing the system's behavior for a sufficiently long time, and then diagnosis can find possible explanations for the given sequence of observations. Otherwise there are scenarios in which it is impossible to tell whether a fault has occurred or not, no matter for how long the system is observed. Naturally, non-diagnosable systems usually indicate that the system should be augmented with additional sensors monitoring it.

A sound software engineering rule for building complex systems is to divide the whole system in smaller and simpler components, each performing a specific task. Moreover, they could be built by different groups of people or in different places. This means that, in general, complex systems are actually collections of simpler components running in parallel. 

In this paper we propose a distributed diagnosability verification with LTL-X model checking based on Petri net unfoldings. We start modeling components as automata, but instead of making a composition of automata we consider the complete system as a Petri Net. Thus, taking the advantage of the compactness of the representation allowed by Petri nets compared to automata. Then, our system is modeled as a collection of components represented as Petri nets and synchronizing on common observable transitions. Also, we remove the assumption that a kind of fault can only occur in a single component (which is usually made in the diagnosability analysis of distributed systems), and allow the same kind of fault to occur in several components (moreover, we allow the same fault to occur in different components; an example of such a fault could be an electricity black out, which can happen in any component independently). 

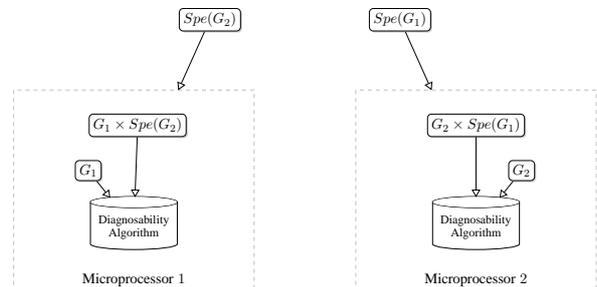
\begin{figure}[h]
  \scalebox{.5}{\newcommand{\An}{\mathcal{A}}
\newcommand{\K}{\mathcal{K}}
\newcommand{\M}{\mathcal{M}}
\newcommand{\T}{\mathcal{T}}

\tikzset{
  sshadow/.style={opacity=.25, shadow xshift=0.05, shadow yshift=-0.06},
}

\def\schemel[#1,#2,#3,#4,#5,#6]#7{ %
  \node[draw, diamond, shape aspect=#3, rotate=#2, minimum size=#1, %
  bottom color=green!55, top color=green!25, color=green!65!black, %
  drop shadow={sshadow,color=green!60!black}, #4] (#5) at #6
  {\textcolor{green!40}{bla}}; %
  \node at #6 {#7};%
}

\def\schemer[#1,#2,#3,#4,#5,#6]#7{ %
  \node[draw, diamond, shape aspect=#3, rotate=#2, minimum size=#1, %
  bottom color=green!65, top color=green!30, color=green!60!black, %
  drop shadow={sshadow,color=green!65!black}, #4] (#5) at #6
  {\textcolor{green!53}{bla}}; %
  \node at #6 {#7}; %
}

\def\tboxl[#1,#2,#3,#4,#5]#6{%
  \node[draw, drop shadow={opacity=.35}, minimum height=#1, minimum width=#2, %
  ] (#4) at #5 {}; %
  \node[anchor=#3,inner sep=2pt] at (#4.#3) {#6};%
}

\def\tboxr[#1,#2,#3,#4,#5]#6{%
  \node[draw, drop shadow={opacity=.35}, minimum height=#1, minimum width=#2, %
  ] (#4) at #5 {}; %
  \node[anchor=#3,inner sep=2pt] at (#4.#3) {#6}; %
}

\def\entity[#1,#2]#3;{
  \node[draw,drop shadow={opacity=.4,shadow xshift=0.04, shadow
    yshift=-0.04},,fill=white,rounded corners=3] (#1) at #2 {#3};
}

\def\isaedge[#1,#2,#3,#4];{ 
  \draw[-triangle 60,color=black!20!black,#4,fill=white] (#1) -- #3
  (#2);  
}

\def\aboxl[#1,#2,#3,#4,#5]#6{%
  \node[draw, cylinder, alias=cyl, shape border rotate=90, aspect=#3, %
  minimum height=#1, minimum width=#2, outer sep=-0.5\pgflinewidth, %
  ] (#4) at #5 {};%
  \node at #5 {#6};%
  \fill [white] let \p1 = ($(cyl.before top)!0.5!(cyl.after top)$), \p2 =
  (cyl.top), \p3 = (cyl.before top), \n1={veclen(\x3-\x1,\y3-\y1)},
  \n2={veclen(\x2-\x1,\y2-\y1)} in (\p1) ellipse (\n1 and \n2); }
    
\def\aboxr[#1,#2,#3,#4,#5]#6{%
  \node[draw, cylinder, alias=cyl, shape border rotate=90, aspect=#3, %
  minimum height=#1, minimum width=#2, outer sep=-0.5\pgflinewidth, %
  ] (#4) at #5 {};%
  \node at #5 {#6};%
  \fill [white] let \p1 = ($(cyl.before top)!0.5!(cyl.after top)$), \p2 =
  (cyl.top), \p3 = (cyl.before top), \n1={veclen(\x3-\x1,\y3-\y1)},
  \n2={veclen(\x2-\x1,\y2-\y1)} in (\p1) ellipse (\n1 and \n2); }

\def\kbbox[#1,#2,#3,#4,#5]#6{
        \draw[dashed] node[draw,color=gray!50,minimum
        height=#1,minimum width=#2] (#4) at #5 {}; 
        \node[anchor=#3,inner sep=2pt] at (#4.#3)  {#6};
}

\def\soledge[#1,#2,#3,#4];{
        \draw[dashed,-latex,#4] (#1) -- #3 (#2);
}

  \begin{tikzpicture}
    \small
    
    \entity[spe1,(2,2)] {\large $Spe(G_2)$};
    \entity[spe2,(7,2)] {\large $Spe(G_1)$};

    \kbbox[150,180,south,k1,(0,-2.5)] {\large Microprocessor $1$};
    \entity[A1,(-1.2,-2)] {\large $G_1$};
    \entity[B1,(0.1,-.8)] {\large $G_1 \times Spe(G_2)$};
    \aboxr[40,65,1.4,a1,(0,-3.5)] {\begin{tabular}{c}Diagnosability\\ Algorithm\end{tabular}};
    \isaedge[A1,a1,,];
    \isaedge[B1,a1,,];
    \isaedge[spe1,k1,,];
        
    \kbbox[150,180,south,k2,(9,-2.5)] {\large Microprocessor $2$};
    \entity[A2,(10.2,-2)] {\large $G_2$};
    \entity[B2,(9,-.8)] {\large $G_2 \times Spe(G_1)$};
    \aboxr[40,65,1.4,a2,(9,-3.5)] {\begin{tabular}{c}Diagnosability\\ Algorithm\end{tabular}};
    \isaedge[A2,a2,,];
    \isaedge[B2,a2,,];
    \isaedge[spe2,k2,,];

  \end{tikzpicture}}
  \caption{Distribution of the diagnosability analysis.}
  \label{fig:motivation}
\end{figure}

We distribute the diagnosability analysis (which is usually done interactively: the information from local components is combined until a global verdict is reached) as it is shown in Figure~\ref{fig:motivation} where $G_1, G_2$ represent two components of the system. Suppose different groups are contracted to build different components of a system. Even if each component is diagnosable, it is not always the case that the resulting system has such property. We propose a framework where each component only shares with the others a fault free version of its own (e.g. the specification of its ideal behavior). Then, each component should not only be diagnosable, but also its interaction with the fault free version of the other components should be diagnosable. We prove that if that is the case then the complete system is diagnosable, resulting in a diagnosability analysis that is distributed.

Finally, we employ the efficient LTL-X model checking based on Petri net unfolding to verify the distributed diagnosability property. Not only we distribute the diagnosability verification but we are employing true concurrency (or partial order) semantics to represent and to check the diagnosability property - which results in important memory savings since executions are considered as partially ordered sets of events rather than sequences.

Our approach extends the distributed diagnosability verification in the framework of automata~\cite{par_diag}; and, it uses existing twin plant method de\-plo\-yed in~\cite{Madalinski2010} where \punf~\cite{punf} is applied to diagnosability verification. 

The paper is organized as follows. First, we show related work in Section~\ref{sec:RelatedWork}. Then, we present the formal model that we use for modeling the system together with basic notions about verification in Section~\ref{sec:DDES}. Section~\ref{sec:diag} introduces the notion of diagnosability and its verification, followed by our main contribution of this paper, the distributed diagnosability analysis, in Section~\ref{sec:DDA}. Finally, we conclude and discuss future work in Section~\ref{sec:CFW}.

\section{Related Work} \label{sec:RelatedWork}

Diagnosability was initially developed in~\cite{Sampath1995} under the setting of discrete event systems. In that paper, necessary and sufficient conditions for testing diagnosability are given. In order to test diagnosability, a special diagnoser is computed, whose complexity of construction is shown to be exponential in the number of states of the original system, and double exponential in the number of faults. Later, in~\cite{twin}, an improvement of this algorithm is presented, where the so-called twin plant method is introduced and shown to have polynomial complexity in the number of states and faults. 
None of the previous methods consider the problem when the system is composed of components working in parallel. An approach to this consideration is addressed in~\cite{Schumann07scalablediagnosability,Debouk_acoordinated,distributeddiag,SchumannH08} where the diagnosability problem is performed by either local diagnosers or twin plants communicating with each other, directly or through a coordinator, and by that means pooling together the observations. \cite{YeDagueValid} shows that, when considering only local observations, diagnosability becomes undecidable when the communication between component is unobservable. An algorithm is proposed to check a sufficient but not necessary condition of diagnosability. However, their results are based on the assumption that a fault can only occur in one of the components, an assumption that cannot always be made.

The state-based twin plant method usually suffers from the combinatorial \emph{state space explosion} problem. That is, even a relatively small system specification can (and often does) yield a very large state space. To alleviate this problem Petri net \emph{unfolding techniques} appear promising. The system is modeled as a Petri net, where each transition is labelled with the performed action. A \emph{finite and complete prefix} of the unfolding gives a compact representation of all the state space. Executions are considered as partially ordered sets of transitions rather than sequences, which often results in memory savings. Since the introduction of the unfolding technique in~\cite{McMillan1992a}, it was improved~\cite{Esparza2002}, parallelized~\cite{Heljanko2002}, and applied to various practical applications such as distributed diagnosis~\cite{Fabre2005a} and LTL-X model checking~\cite{EH-01}. Also, the problem of diagnosability verification based on the twin plant method has been studied in~\cite{Madalinski2010} in the context of parallel LTL-X model checking based on Petri net unfoldings.

\section{Basic Notions} \label{sec:DDES}

\paragraph{Model of the system.}
We consider distributed systems composed by several components that communicate with each other through their shared observable actions, as diagnosability is undecidable when communication is unobservable~\cite{YeDagueValid}. The local model of a component is defined as an automaton $(Q,\Sigma,\delta,q_0)$, where $Q$ is a finite set of states, $\Sigma$ is a finite set of actions, $\delta : Q \times \Sigma \rightarrow Q$ is the transition function and $q_0 \in Q$ is the initial state.

In diagnosability analysis, some of the actions of $\Sigma$ are observable while the rest are unobservable. Thus, the set of actions $\Sigma$ is partitioned as $\Sigma=\Sigma_o\uplus \Sigma_u$ where $\Sigma_o$ represents the observable actions and $\Sigma_u$ the unobservable ones. The faults to diagnose are considered unobservable, i.e. $\Sigma_F \subseteq \Sigma_u$, because faults that are observable can be easily diagnosed.

As usual in diagnosability analysis, we made the following assumptions about our systems.

\begin{assum}
  We only consider (live) systems, where there is a transition defined at each state, i.e. the system cannot reach a point at which no action is possible.
\end{assum}

\begin{assum}
  The system does not contain cycles of unobservable actions.
\end{assum}

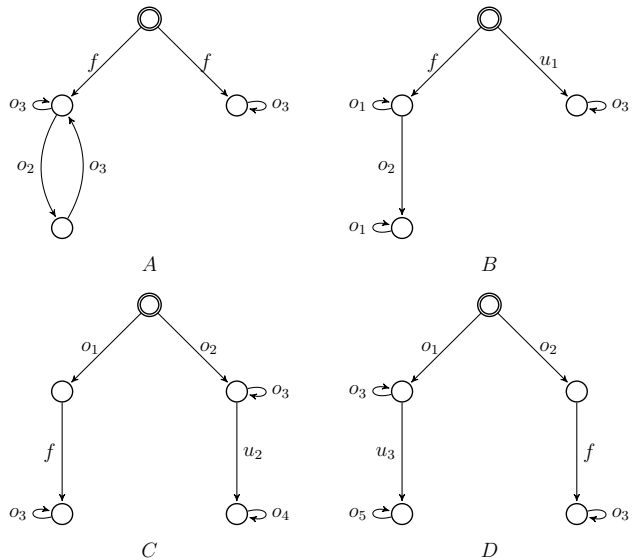
\begin{figure}
   \centering
   \subfigure{\scalebox{.65}{\begin{tikzpicture}[->,>=stealth',shorten >=1pt,auto,node distance=2.5cm,semithick]
\tikzstyle{every state}=[draw]

  \node[accepting,state] 		(A)                    {};
  \node[state]         	(B) [below left of=A] {};
  \node[state]         	(C) [below right of=A] {};
  \node[state]         	(D) [below of=B] {};

  \path	(A)	edge [left]		node {\Large $f$}		(B)
            	(A)	edge	 [right]	node {\Large $f$}		(C)
	         (B)	edge	 [bend right, left]		node {\Large $o_2$}		(D)
		(B)	edge	 [loop left]	node {\Large $o_3$}		(B)
		(D)	edge	 [bend right,right]	node {\Large $o_3$}		(B)
		(C)	edge	 [loop right]	node {\Large $o_3$}		(C);
		
\node[] (name) at (0,-5) {\Large $A$};					

\end{tikzpicture}}} \hspace{4mm}
   \subfigure{\scalebox{.65}{\begin{tikzpicture}[->,>=stealth',shorten >=1pt,auto,node distance=2.5cm,semithick]
\tikzstyle{every state}=[draw]

  \node[accepting,state] 		(A)                    {};
  \node[state]         	(B) [below left of=A] {};
  \node[state]         	(C) [below right of=A] {};
  \node[state]         	(D) [below of=B] {};

\draw[->]	(A)	edge [left]		node {\Large{\bf $f$}}	(B);
\draw[->]	(B)	edge	 [loop left]	node {\Large $o_1$}		(B);
\draw[->]	(A)	edge	 [right]	node {\Large $u_1$}		(C);
\draw[->]	(B)	edge	 [left]		node {\Large $o_2$}		(D);
\draw[->]	(D)	edge	 [loop left]	node {\Large $o_1$}		(D);
\draw[->]	(C)	edge	 [loop right]node {\Large $o_3$}	(C);
		
\node[] (name) at (0,-5) {\Large $B$};							

\end{tikzpicture}}}
   \subfigure{\scalebox{.65}{\begin{tikzpicture}[->,>=stealth',shorten >=1pt,auto,node distance=2.5cm,semithick]
\tikzstyle{every state}=[draw]

  \node[accepting,state] 		(A)                    {};
  \node[state]         	(B) [below left of=A] {};
  \node[state]         	(C) [below right of=A] {};
  \node[state]         	(D) [below of=B] {};
  \node[state]         	(E) [below of=C] {};

  \path	(A)	edge [left]		node {\Large $o_1$}		(B)
            	(A)	edge	 [right]		node {\Large $o_2$}		(C)
	         (C)	edge	 [loop right]	node {\Large $o_3$}		(C)
		(B)	edge	 [left]		node {\Large $f$}		(D)
		(D)	edge	 [loop left]	node {\Large $o_3$}		(D)	
		(C)	edge	 [right]		node {\Large $u_2$}		(E)
		(E)	edge	 [loop right]	node {\Large $o_4$}		(E);

\node[] (name) at (0,-5) {\Large $C$};			

\end{tikzpicture}}} \hspace{4mm}
   \subfigure{\scalebox{.65}{\begin{tikzpicture}[->,>=stealth',shorten >=1pt,auto,node distance=2.5cm,semithick]
\tikzstyle{every state}=[draw]

  \node[accepting,state] 		(A)                    {};
  \node[state]         	(B) [below right of=A] {};
  \node[state]         	(C) [below left of=A] {};
  \node[state]         	(D) [below of=B] {};
  \node[state]         	(E) [below of=C] {};

  \path	(A)	edge [right]		node {\Large $o_2$}		(B)
            	(A)	edge	 [left]		node {\Large $o_1$}		(C)
	         (C)	edge	 [loop left]	node {\Large $o_3$}		(C)
		(B)	edge	 [right]		node {\Large $f$}		(D)
		(D)	edge	 [loop right]	node {\Large $o_3$}		(D)	
		(C)	edge	 [left]		node {\Large $u_3$}		(E)
		(E)	edge	 [loop left]	node {\Large $o_5$}		(E);

\node[] (name) at (0,-5) {\Large $D$};			

\end{tikzpicture}}}
   \caption{Components specifications as automata}
   \label{fig:lts}
 \end{figure}

Figure~\ref{fig:lts} shows four components modeled by automata $A,B,C$ and $D$ where $o_1,o_2,o_3,o_4,o_5 \in \Sigma_o$ and $u_1,u_2,u_3 \in \Sigma_u$. The special action $f \in \Sigma_F$ is the fault to be diagnosed.

The joint behavior of the system can be represented by a safe labelled Petri net. A \emph{labelled net} is a tuple $\onlynetn=(\places,\trans,\flow,\lambda)$ where \emph{(i)}  $\places\ne\emptyset$ is a set of \emph{places}, \emph{(ii)}  $\trans\ne\emptyset$ is a set of \emph{transitions} such that $\places\cap\trans=\emptyset$, \emph{(iii)}  $\flow\subseteq(\places\times\trans)\cup(\trans\times\places)$ is a set of \emph{flow arcs}, \emph{(iv)} $\lambda: T \rightarrow \Sigma$ is a \emph{labelling} function. A \emph{marking} is a subset $\markn$ of places, i.e. $\markn\subseteq \places$. A \emph{labelled Petri net} is a tuple $\netn=(\places,\trans,\flow,\lambda, \markn_0)$, where \emph{(i)} $(\places,\trans,\flow,\lambda)$ is a finite labelled net, and \emph{(ii)} $\markn_0 \subseteq \places$ is an \emph{initial marking}. Elements of $\places\cup\trans$ are called the \emph{nodes} of $\netn$. For a transition  $\transn\in\trans$, we call $\preset{\transn}=\{\placen\ |\ (\placen,\transn)\in\flow\}$ the \emph{preset} of $t$, and $\postset{t}=\{p\ |\ (t,p)\in\flow\}$ the \emph{postset} of $\transn$. In figures, we represent, as usual, places by empty circles; transitions by squares; $\flow$ by arrows; and the marking of a place $\placen$ by  black tokens in $\placen$.  A transition $\transn$ is \emph{enabled} in marking $\markn$, written $\markn\move{\transn}$, if $\forall \placen\in\preset{\transn},\; \markn(\placen) > 0$. This enabled transition can \emph{fire}, resulting in a new marking $\markn'=\markn-{\preset{\transn}}+{\postset{\transn}}$. This firing relation is denoted by $\markn\move{\transn}\markn'$. A \emph{run} is a sequence $\rho=M_0t_oM_1t_1,\dots t_{n-1}M_n$ such that $M_0 \move{t_0} M_1 \move{t_1} \dots \move{t_{n-1}} M_n$ and $\sigma= \lambda(t_0)\lambda(t_1)\dots\lambda(t_{n-1})$ is its associated \emph{trace}, i.e $\trace{\rho}=\sigma$. A marking $\markn$ is \emph{reachable} if there exists a run from $M_0$ to $M$. The set of markings reachable from $\markn_0$ is denoted $\reach(\markn_0)$. 

As our systems are live, we only consider infinite traces where the infinite repetition of action $a$ is denoted by $\widehat a$. The set of runs and traces are denoted by $\runs \netn$ and $\inftraces \netn$ respectively. As only some actions are observable, the observable projection of a trace is defined as
\begin{eqnarray*}
  \obs \sigma & = &
  \left \{
  \begin{array}{l l l}
    \epsilon & \text{if } \sigma = \epsilon \\
    a\!\cdot\! \obs{\sigma'} & \text{if } \sigma = a\!\cdot\!\sigma' \land a\!\in\!\Sigma_o \\
    \obs{\sigma'} & \text{if } \sigma = a\!\cdot\!\sigma' \land a\!\not \in\!\Sigma_o \\
  \end{array}
  \right.
\end{eqnarray*}

The translation from an automaton $A$ to a labelled Petri net $\netn_A$ is immediate: \emph{(i)} places are the states of the automaton, i.e. $P=Q$; \emph{(ii)} for every transition $(s_i,a,s_i') \in \delta$ we add $t$ to $T$ and set $\preset t = \{ s_i\}, \postset t = \{ s_i' \}$ and $\lambda(t)=a$; \emph{(iii)} the initial state is the only place marked initially, i.e. $M_0 = \{ q_0 \}$. 

The joint behavior of a system composed of automata $\{A_1, \dots, A_n\}$ is modeled by $\netn_{A_1} \times \dots \times \netn_{A_n}$ where $\times$ represents the product of labelled nets defined in~\cite{prod_nets} synchronizing on shared observable transitions. Product of nets prevents us from the state explosion problem that usually arises in product of automata.

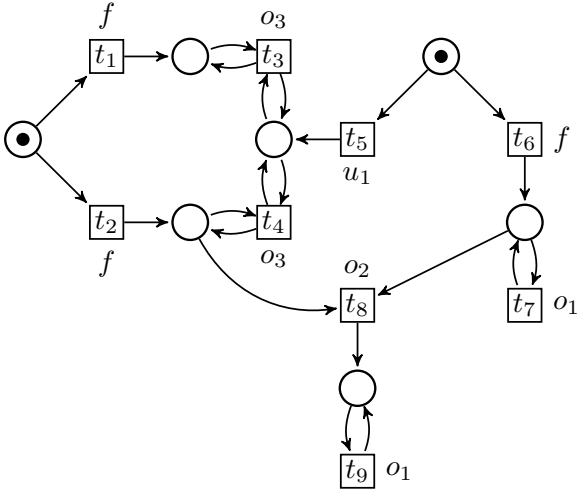
\begin{figure}[h]
  \centering
  \subfigure{\scalebox{1.1}{\begin{tikzpicture}[->,>=stealth',shorten >=1pt,auto,node distance=2.5cm,semithick]
\tikzstyle{every state}=[draw]

\node[state] (p1) {};
\node[state] (p2) at (2,1) {};
\node[state] (p3) at (2,-1) {};
\node[state] (p4) at (3,0) {};
\node[state] (p5) at (5,1) {};
\node[state] (p6) at (6,-1) {};
\node[state] (p8) at (4,-3) {};

\node[transition,label=above:$f$] (t1) at (1,1) {$t_1$};
\node[transition,label=below:$f$] (t2) at (1,-1) {$t_2$};
\node[transition,label=above:$o_3$] (t3) at (3,1) {$t_3$};
\node[transition,label=below:$o_3$] (t4) at (3,-1) {$t_4$};
\node[transition,label=below:$u_1$] (t5) at (4,0) {$t_5$};
\node[transition,label=right:$f$] (t6) at (6,0) {$t_6$};
\node[transition,label=right:$o_1$] (t7) at (6,-2) {$t_7$};
\node[transition,label=above:$o_2$] (t8) at (4,-2) {$t_8$};
\node[transition,label=right:$o_1$] (t9) at (4,-4) {$t_9$};

\node[token] at (p1) {};
\node[token] at (p5) {};

\node[] (name) at (3.5,-5) {$\netn_1 = \netn_A \times \netn_B$};

\draw[->] (p1) to (t1);
\draw[->] (t1) to (p2);
\draw[<-, bend right=20] (t3) to (p2);
\draw[->, bend left=20] (t3) to (p2);
\draw[<-, bend right=20] (t3) to (p4);
\draw[->, bend left=20] (t3) to (p4);
\draw[->] (p1) to (t2);
\draw[->] (t2) to (p3);
\draw[<-, bend right=20] (t4) to (p3);
\draw[->, bend left=20] (t4) to (p3);
\draw[<-, bend right=20] (t4) to (p4);
\draw[->, bend left=20] (t4) to (p4);
\draw[->,bend right=35] (p3) to (t8);
\draw[->] (p6) to (t8);
\draw[->] (t8) to (p8);
\draw[<-, bend left=20] (p8) to (t9);
\draw[->, bend right=20] (p8) to (t9);

\draw[->] (p5) to (t5);
\draw[->] (t5) to (p4);
\draw[->] (p5) to (t6);
\draw[->] (t6) to (p6);
\draw[<-, bend right=20] (p6) to (t7);
\draw[->, bend left=20] (p6) to (t7);

\end{tikzpicture}}} 
  \caption{Automata $\{ A,B \}$ represented as Petri nets}
  \label{fig:pn1}
\end{figure}

\begin{figure}[h]
  \centering
  \subfigure{\scalebox{1}{\begin{tikzpicture}[->,>=stealth',shorten >=1pt,auto,node distance=2.5cm,semithick]
\tikzstyle{every state}=[draw]

\node[state] (p1) at (.5,0){};
\node[state] (p2) at (1.5,0) {};
\node[state] (p3) at (2,1) {};
\node[state] (p4) at (2,-1) {};
\node[state] (p5) at (4,3) {};
\node[state] (p6) at (4,1) {};
\node[state] (p7) at (4,-1) {};
\node[state] (p8) at (4,-3) {};
\node[state] (p9) at (6,3) {};
\node[state] (p10) at (6,-3) {};

\node[transition,label=above:$o_1$] (t1) at (1,2) {$t_1$};
\node[transition,label=below:$f$] (t2) at (3,1) {$t_2$};
\node[transition,label=left:$o_3$] (t3) at (4,2) {$t_3$};
\node[transition,label=above:$u_3$] (t4) at (5,3) {$t_4$};
\node[transition,label=right:$o_5$] (t5) at (6,2) {$t_5$};
\node[transition,label=below:$o_2$] (t6) at (1,-2) {$t_6$};
\node[transition,label=above:$f$] (t7) at (3,-1) {$t_7$};
\node[transition,label=left:$o_3$] (t8) at (4,-2) {$t_8$};
\node[transition,label=below:$u_2$] (t9) at (5,-3) {$t_9$};
\node[transition,label=right:$o_4$] (t10) at (6,-2) {$t_{10}$};

\node[token] at (p1) {};
\node[token] at (p2) {};

\node[] (name) at (3.5,-4) {$\netn_2 = \netn_C \times \netn_D$};

\draw[->] (p1) to (t1);
\draw[->] (p2) to (t1);
\draw[->] (t1) to (p3);
\draw[->] (t1) to (p5);
\draw[->] (p3) to (t2);
\draw[->] (t2) to (p6);
\draw[<-, bend right = 20] (p6) to (t3);
\draw[->, bend left = 20] (p6) to (t3);
\draw[<-, bend right = 20] (p5) to (t3);
\draw[->, bend left = 20] (p5) to (t3);
\draw[->] (p5) to (t4);
\draw[->] (t4) to (p9);
\draw[<-, bend right = 20] (p9) to (t5);
\draw[->, bend left = 20] (p9) to (t5);

\draw[->] (p1) to (t6);
\draw[->] (p2) to (t6);
\draw[->] (t6) to (p4);
\draw[->] (t6) to (p8);
\draw[->] (p4) to (t7);
\draw[->] (t7) to (p7);
\draw[<-, bend right = 20] (p7) to (t8);
\draw[->, bend left = 20] (p7) to (t8);
\draw[<-, bend right = 20] (p8) to (t8);
\draw[->, bend left = 20] (p8) to (t8);
\draw[->] (p8) to (t9);
\draw[->] (t9) to (p10);
\draw[<-, bend right = 20] (p10) to (t10);
\draw[->, bend left = 20] (p10) to (t10);

\end{tikzpicture}}}
  \caption{Automata $\{ C,D \}$ represented as Petri nets}
  \label{fig:pn2}
\end{figure}

The joint behavior of $\{A, B\}$ and $\{C, D\}$ from Figure~\ref{fig:lts} can be modeled by the corresponding Petri nets $\netn_1= \netn_A \times \netn_B$ and $\netn_2 = \netn_C \times \netn_D$ of Figure~\ref{fig:pn1} and Figure~\ref{fig:pn2}. 

Consider the automata $\{A_1,\dots, A_n\}$ and its corresponding net $\netn$. The projection of a run on component $i$ is given by
\begin{eqnarray*}
  P_i((q^1, \dots, q^n)) & \hspace{-0.2cm}=\hspace{-0.2cm} & q^i \\
  P_i((q^1, \dots, q^n)\!\cdot\!t\!\cdot\!\rho') & \hspace{-0.2cm}=\hspace{-0.2cm} & 
  \left \{
  \begin{array}{lcl}
  q^i\!\cdot\!t\!\cdot\!P_i(\rho') & &\hspace{-0.2cm}\text{if } \exists\ \delta^i(q^i, \lambda(t)) \\
  P_i(\rho') & & \text{otherwise}
  \end{array}
  \right.
\end{eqnarray*}

For $\sigma \in \inftraces \netn$, we say that $\sigma_i$ is its projection on component $i$, denoted $\proj i \sigma = \sigma_i$, if and only if $\exists \rho \in \camino \sigma: \trace{\proj i \rho}=\sigma_i$.

\begin{exmp}
  Consider $\sigma = o_1fo_3u_3\widehat{o}_5 \in \inftraces{\netn_2}$. Its projection on components $C$ and $D$ are given by $\proj C \sigma = o_1fo_3$ and $\proj D \sigma = o_1o_3u_3\widehat{o}_5$. These projections are traces of the corresponding components $C$ and $D$ from Figure~\ref{fig:lts}. Note that projections of an infinite trace from the net can be finite in one component.
\end{exmp}

As the projection operator only erases actions in a trace, it is easy to see that every fault belonging to a trace of a component, also belongs to the trace of the net as it is shown by the following result.

\begin{prop} 
  Let $\netn = \netn_{A_1}\times \dots \times \netn_{A_n}$, then for every $\sigma \in \inftraces \netn$ with $\proj i \sigma = \sigma_i$, if $f \in \sigma_i$ then $\imp f \in \sigma$.
  \label{prop:fault_proj}
\end{prop}

When two traces of the net have the same observability and we project them on the same component, the resulting projections also have the same observability. This result is captured by the following proposition.

\begin{prop}
  Consider the net $\netn = \netn_{A_1}\times \dots \times \netn_{A_n}$ and $\sigma, \alpha \in \inftraces \netn$ with $\proj i \sigma = \sigma_i$ and $\proj i \alpha = \alpha_i$, we have $\obs \sigma = \obs \alpha$ implies $\obs{\sigma_i} = \obs{\alpha_i}$
  \label{prop:obs}.
\end{prop}

Note that this result only holds because the product of nets synchronize on the set of shared actions.

\paragraph{Unfolding prefixes.}

The \emph{unfolding} of a Petri net $\netn$ is a (potentially infinite) acyclic net that can be obtained by starting from the initial marking of $\netn$ and successively firing its transitions, as follows: (a) for each new firing a fresh transition (called an \emph{event}) is generated; (b) for each newly produced token a fresh place (called a \emph{condition}) is generated. Due to its structural properties, the reachable markings of $\netn$ can be represented using \emph{configurations} of the unfolding. Intuitively, a configuration is a finite partially ordered execution, i.e. an execution where the order of firing of concurrent events is not important.

The unfolding is infinite whenever $\netn$ has an infinite execution; however, if $\netn$ is bounded (and thus has finitely many reachable states) then the unfolding eventually starts to repeat itself and can be truncated (by identifying a set of \emph{cut-off events}) without loss of information, yielding a \emph{finite and complete prefix.} 

\paragraph{LTL-X and \buchi automata.}

\emph{Linear time temporal logic} (LTL)~\cite{P-77} is a logic allowing to specify the properties of computations, and LTL-X is the fragment of LTL obtained by removing the \emph{next-state} modality. LTL-X plays a prominent role in formal verification.

Deciding whether all computations of system $S$ satisfy $\varphi$ is equivalent to deciding whether some computation of $S$ satisfies $\neg\varphi$~\cite{VW-86}. Formula $\neg\varphi$ is converted into a \emph{\buchi automaton} $A_{\neg\varphi}$ accepting the computations satisfying $\neg\varphi$~\cite{GO-01}. Then, $S$ and $\A$ are synchronized in such a way that the language of the resulting \buchi automaton $S\times\A$ is the intersection of the language of $\A$ and the set of all the possible computations of $S$. Hence, in this way one can reduce the original verification problem to checking if the language accepted by the \buchi automaton $S\times\A$ is empty, which can be efficiently solved.

\paragraph{Unfolding based LTL-X model checking.}

In~\cite{EH-01} an efficient approach to model checking LTL-X properties of Petri nets based on unfolding prefixes was proposed. Its main differences from the automata-based approach outlined above are the following. The \buchi automaton $\A$ for the LTL-X property $\varphi$ is translated into a Petri net $\N$, called \emph{\buchi net} (simply by replacing the automata states by places and automata transitions by transitions). Then its synchronization with the Petri net model of system $S$ is performed at the level of Petri nets rather than reachability graphs, resulting in another \buchi net. The synchronization is defined such that the concurrency present in $S$ is preserved as much as possible, which is important for the subsequent unfolding. Then the resulting synchronization is unfolded, and
the cut-off events are defined such that the resulting finite and complete prefix can be viewed as a tableau proof, from which it is easy either to conclude that the property holds or to find a trace of $S$ violating the property. This approach can significantly outperform methods based on explicit construction of reachability graphs in case of highly concurrent systems.

\section{Diagnosability Analysis} \label{sec:diag}

We present now the notion of diagnosability. Informally, a fault $f \in \Sigma_F$ is diagnosable if it is possible to detect, within a finite delay, occurrences of such a fault using the record of observed actions. In other words, a fault is not diagnosable if there exist two infinite runs from the initial state with the same infinite sequence of observable actions but only one of them contains the fault.

\begin{defn}
  A fault $f$ is diagnosable in $\netn$ iff $\forall \sigma, \alpha \in \inftraces \netn: \obs \sigma = \obs \alpha$ and $f \in \sigma$ implies $f \in \alpha$. $\netn$ is diagnosable, denoted by $\diag \netn$, if and only if every fault $f \in \Sigma_F$ is diagnosable.
\end{defn}

As automata can be seen as nets with no concurrency and diagnosability is a property that consider (sequential) runs, the above definition can also be applied for automata.

\begin{prop}
 Consider the automaton $A$ and its corresponding net $\netn_A$, we have $\diag A \Leftrightarrow \diag{\netn_A}$.
\end{prop}

\begin{exmp}
  Consider the components $A$ and $B$ from Figure~\ref{fig:lts}. The only pair of traces in $A$ with the same observability are of the form $f\widehat{o}_3$ (one for each branch from the initial state). As both traces contain the fault $f$, system $A$ is diagnosable. In the case of $B$, each observable trace corresponds to a unique run, therefore $B$ is diagnosable. Now, consider the net $\netn_1$ from Figure~\ref{fig:pn1} and Figure~\ref{fig:pn2}, we can see that every trace contains a fault, therefore $\netn_1$ is diagnosable. For net $\netn_2$ from Figure~\ref{fig:pn2} we have two traces, $o_2u_2\widehat{o}_4$ and $o_2fu_2\widehat{o}_4$ that have the same observability, but one of them contains a fault and the other does not, therefore $\netn_2$ is not diagnosable.
  \label{ex:one}
\end{exmp}

The product of automata is usually much bigger than the product of their corresponding nets as every possible interleaving is constructed, however there is an isomorphism between their runs~\cite{runs_prod} and we have the following result.

\begin{prop}
  Let $\{A_1, \dots, A_n\}$ be a set of automata, then $\diag{A_1 \times \dots \times A_n} \Leftrightarrow \diag{\netn_{A_1} \times \dots \times \netn_{A_n}}$.
\end{prop}

We can now exploit the concurrency of the system and analyze its diagnosability by the verification of Petri nets. 

\paragraph{LTL-X model checking for non-diagnosability.}

The diagnosability property is verified using LTL-X model checking based on Petri net unfoldings~\cite{Madalinski2010}.  The \emph{verifier} \veri is built with respect to a fault $f$ by synchronizing two replicas of $\netn$ on the observable transitions. Note that for efficiency reasons one replica does not consider the fault. 

Intuitively, the two replicas are put side-by-side, and then each observable transition in the first replica is fused with each transition in the second replica that has the same label (each fusion produces a new transition, and the original observable transitions are removed). One can see that there is a one-to-one correspondence between the traces of \veri and pairs of traces of $\netn$ with the same projections on the set of observable actions.

As explained above, given the verifier \veri, checking the complement $\ndiagltl$ of the diagnosability property can be reduced to checking the existence of an infinite trace of \veri containing an occurrence of $f$, in LTL-X it can be expressed as $\ndiagltl\DEF\eventually f$, where $\eventually$ is the modality \emph{eventually}.

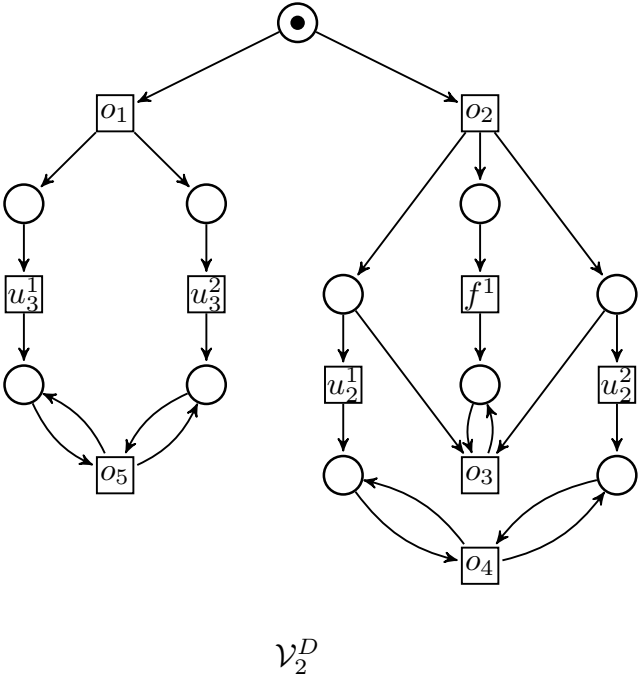
\begin{figure}[h]
   \centering
   \scalebox{1.2}{\begin{tikzpicture}[->,>=stealth',shorten >=1pt,auto,node distance=2.5cm,semithick]
\tikzstyle{every state}=[draw]

\node[state] (p1) {};

\node[transition, xshift=1cm] (t1) at (-3,-1) {$o_1$};
\node[state, xshift=1cm] (p2) at (-4,-2) {};
\node[state, xshift=1cm] (p4) at (-2,-2) {};
\node[transition, xshift=1cm] (t4) at (-4,-3) {$u^1_3$};
\node[transition, xshift=1cm] (t4') at (-2,-3) {$u^2_3$};
\node[state, xshift=1cm] (p5) at (-4,-4) {};
\node[state, xshift=1cm] (p7) at (-2,-4) {};
\node[transition, xshift=1cm] (t5) at (-3,-5) {$o_5$};

\node[transition, xshift=-1cm] (t6) at (3,-1) {$o_2$};
\node[state, xshift=-1cm] (p8) at (3,-2) {};
\node[state, xshift=-1cm] (p9) at (1.5,-3) {};
\node[state, xshift=-1cm] (p10) at (4.5,-3) {};
\node[transition, xshift=-1cm] (t7) at (3,-3) {$f^1$};
\node[state, xshift=-1cm] (p11) at (3,-4) {};
\node[transition, xshift=-1cm] (t8) at (3,-5) {$o_3$};
\node[state, xshift=-1cm] (p12) at (1.5,-5) {};
\node[state, xshift=-1cm] (p13) at (4.5,-5) {};
\node[transition, xshift=-1cm] (t10) at (3,-6) {$o_4$};
\node[transition, xshift=-1cm] (t9) at (1.5,-4) {$u^1_2$};
\node[transition, xshift=-1cm] (t9') at (4.5,-4) {$u^2_2$};

\node[token] at (p1) {};

\node[] (name) at (0,-7) {$\veri^D_2$};

\draw[->] (p1) to (t1);
\draw[->] (t1) to (p2);
\draw[->] (t1) to (p4);
\draw[->] (p2) to (t4);
\draw[->] (p4) to (t4');
\draw[->] (t4) to (p5);
\draw[->] (t4') to (p7);
\draw[->, bend right=20] (p5) to (t5);
\draw[->, bend right=20] (p7) to (t5);
\draw[<-, bend left=20] (p5) to (t5);
\draw[<-, bend left=20] (p7) to (t5);

\draw[->] (p1) to (t6);
\draw[->] (t6) to (p8);
\draw[->] (t6) to (p9);
\draw[->] (t6) to (p10);
\draw[->] (p8) to (t7);
\draw[->] (p9) to (t8);
\draw[->] (p10) to (t8);
\draw[->] (t7) to (p11);
\draw[->, bend right=20] (p11) to (t8);
\draw[<-, bend left=20] (p11) to (t8);
\draw[->] (p9) to (t9);
\draw[->] (p10) to (t9');
\draw[->] (t9) to (p12);
\draw[->] (t9') to (p13);
\draw[->, bend right=20] (p12) to (t10);
\draw[<-, bend left=20] (p12) to (t10);
\draw[->, bend right=20] (p13) to (t10);
\draw[<-, bend left=20] (p13) to (t10);

\end{tikzpicture}}
   \caption{Verifier of $\netn^D_2$}
   \label{fig:ver}
 \end{figure}

\begin{exmp} 
  The verifier $\veri^D_2$ of the net $\netn^D_2$ (presented in the next section, see Figure~\ref{fig:pn_reduced}) is depicted in Figure~\ref{fig:ver}. The superscript is used to distinguish nodes belonging 
to each copy of ${\netn^D_2}$, e.g. there are two copies of $u_2$ in $\veri^D_2$, $u_2^1$ and $u_2^2$; the fusion transitions do not have superscripts: they are considered `common'. The infinite trace of $\veri^D_2 : o_2 f^1 u_2^1 u_2^2 \widehat{o}_4$ satisfies  the \ndiagltl\ property. This trace of $\veri^D_2$ corresponds to the pair of traces $o_2 f u_2 \widehat{o}_4$ and  $o_2 u_2 \widehat{o}_4$ of $\netn^D_2$, constituting a witness of diagnosability violation.
  \label{ex:veri}
\end{exmp}

\section{Distributing the Diagnosability Analysis} \label{sec:DDA}

In this section we present a method that allows to decide the diagnosability of a distributed system in terms of the diagnosability of each faulty component interacting with fault free versions of the remaining ones. These diagnosability analyses can be done in parallel.

For testing the diagnosability of a fault $f \in \Sigma_F$ in a net $\netn = \netn_{A_1}\times \dots \times \netn_{A_n}$, we consider a component $i$ and compose it with fault free versions of the others, we denote such net as $\netn^i$. These fault free versions may be taken as the specification of each component, when provided, or can be computed by removing the fault $f$ in the net of such component using Algorithm~\ref{algo} and considering it as the correct behavior of the system.

\begin{algorithm}[h]
\caption{}
  \begin{algorithmic}[1] \label{algo1}
    \REQUIRE A Petri net $\netn = (P, T, F, M_0, \lambda)$
    \ENSURE A $f$-fault free version of $\netn$
      \STATE $P' := M_0 $ , $T' := \emptyset$ , $P := P \setminus M_0$
      \WHILE{$\exists t \in T\backslash T': \preset t \subseteq P'$}
        \IF{$\lambda(t) \not = f$}
          \STATE $P' := P' \cup \preset t$
          \STATE $T' := T' \cup \{ t \}$
        \ENDIF
        \STATE $T := T \setminus \{ t \}$
      \ENDWHILE
    \STATE $F' := F \cap ((P' \times T') \cup (T' \times P'))$
    \STATE $\lambda' := \lambda_{\mid T'}$
    \RETURN $\netn^f = (P',T', F', M_0,\lambda')$
  \end{algorithmic}
  \label{algo}
\end{algorithm}

 \begin{figure}[h]
   \centering
   \subfigure{\scalebox{.7}{\begin{tikzpicture}[->,>=stealth',shorten >=1pt,auto,node distance=2.5cm,semithick]
\tikzstyle{every state}=[draw]

  \node[state] 		(A)                    {};
		
\node[] (name) at (0,-5) {\large $\netn_A^f$};					

\end{tikzpicture}}} \hspace{4mm}
   \subfigure{\scalebox{.7}{\begin{tikzpicture}[->,>=stealth',shorten >=1pt,auto,node distance=2.5cm,semithick]
\tikzstyle{every state}=[draw]

  \node[state] (A) {};
  \node[state] (C) at (0,-2) {};

\node[transition] (t1) at (0,-1) {$u_1$}; 
\node[transition] (t2) at (1,-2) {$o_3$}; 

\draw[->] (A) to (t1);
\draw[->] (t1) to (C);
\draw[->, bend right = 20] (C) to (t2);
\draw[->, bend right = 20] (t2) to (C);

\node[] (name) at (0,-5) {\large $\netn_B^f$};							

\end{tikzpicture}}}\hspace{4mm}
   \subfigure{\scalebox{.7}{\begin{tikzpicture}[->,>=stealth',shorten >=1pt,auto,node distance=2.5cm,semithick]
\tikzstyle{every state}=[draw]

  \node[state] (A)                    {};
  \node[state] (B) at (-1,-2) {};
  \node[state] (C) at (1,-2) {};
  \node[state] (E) at (1,-4) {};

\node[transition] (t1) at (-1,-1) {$o_1$};
\node[transition] (t2) at (1,-1) {$o_2$};
\node[transition] (t3) at (2,-2) {$o_3$};
\node[transition] (t4) at (1,-3) {$u_2$};
\node[transition] (t5) at (2,-4) {$o_4$};

\draw[->] (A) to (t1);
\draw[->] (t1) to (B);
\draw[->] (A) to (t2);
\draw[->] (t2) to (C);
\draw[->, bend right = 20] (C) to (t3);
\draw[->, bend right = 20] (t3) to (C);
\draw[->] (C) to (t4);
\draw[->] (t4) to (E);
\draw[->, bend right = 20] (E) to (t5);
\draw[->, bend right = 20] (t5) to (E);

\node[] (name) at (0,-5) {\large $\netn_C^f$};			

\end{tikzpicture}}}\hspace{4mm}
   \subfigure{\scalebox{.7}{\begin{tikzpicture}[->,>=stealth',shorten >=1pt,auto,node distance=2.5cm,semithick]
\tikzstyle{every state}=[draw]

  \node[state] (A)                    {};
  \node[state] (B) at (-1,-2) {};
  \node[state] (C) at (1,-2) {};
  \node[state] (E) at (-1,-4) {};

\node[transition] (t1) at (-1,-1) {$o_1$};
\node[transition] (t2) at (1,-1) {$o_2$};
\node[transition] (t3) at (-2,-2) {$o_3$};
\node[transition] (t4) at (-1,-3) {$u_3$};
\node[transition] (t5) at (-2,-4) {$o_5$};

\draw[->] (A) to (t1);
\draw[->] (t1) to (B);
\draw[->] (A) to (t2);
\draw[->] (t2) to (C);
\draw[->, bend right = 20] (B) to (t3);
\draw[->, bend right = 20] (t3) to (B);
\draw[->] (B) to (t4);
\draw[->] (t4) to (E);
\draw[->, bend right = 20] (E) to (t5);
\draw[->, bend right = 20] (t5) to (E);

\node[] (name) at (0,-5) {\large $\netn_D^f$};			

\end{tikzpicture}}}
   \caption{Components after removing their faults}
   \label{fig:reduced}
 \end{figure}
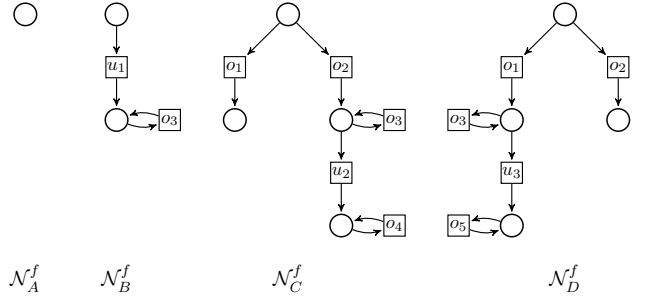

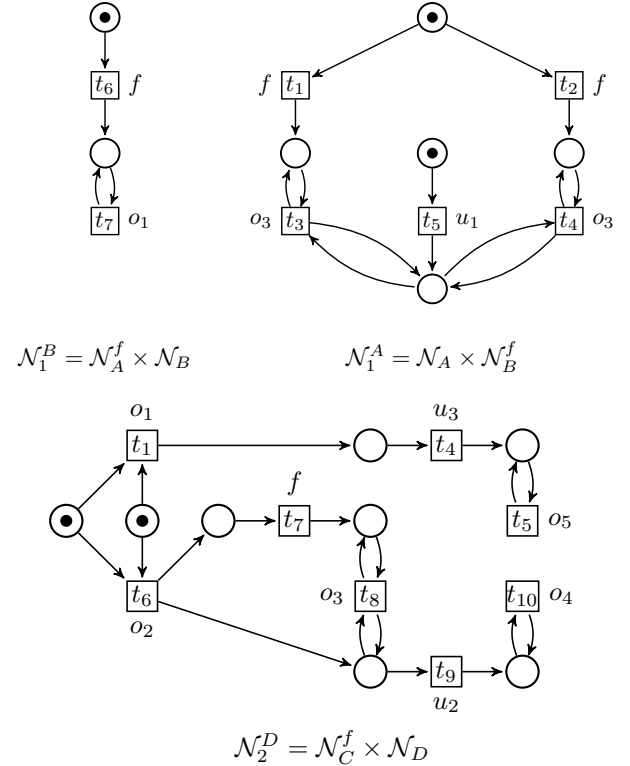
\begin{figure}[h]
   \centering
   \subfigure{\scalebox{.9}{\begin{tikzpicture}[->,>=stealth',shorten >=1pt,auto,node distance=2.5cm,semithick]
\tikzstyle{every state}=[draw]

\node[state] (p5) {};
\node[state] (p6) at (0,-2) {};

\node[transition,label=right:$f$] (t6) at (0,-1) {$t_6$};
\node[transition,label=right:$o_1$] (t7) at (0,-3) {$t_7$};

\node[token] at (p5) {};

\node[] (name) at (0,-5) {$\netn_1^B = \netn_A^f \times \netn_B$};

\draw[->] (p5) to (t6);
\draw[->] (t6) to (p6);
\draw[<-, bend right = 20] (p6) to (t7);
\draw[->, bend left = 20] (p6) to (t7);

\end{tikzpicture}}} \hspace{3mm}
   \subfigure{\scalebox{.9}{\begin{tikzpicture}[->,>=stealth',shorten >=1pt,auto,node distance=2.5cm,semithick]
\tikzstyle{every state}=[draw]

\node[state] (p1) {};
\node[state] (p2) at (-2,-2) {};
\node[state] (p3) at (2,-2) {};
\node[state] (p4) at (0,-4) {};
\node[state] (p5) at (0,-2) {};

\node[transition,label=left:$f$] (t1) at (-2,-1) {$t_1$};
\node[transition,label=right:$f$] (t2) at (2,-1) {$t_2$};
\node[transition,label=left:$o_3$] (t3) at (-2,-3) {$t_3$};
\node[transition,label=right:$o_3$] (t4) at (2,-3) {$t_4$};
\node[transition,label=right:$u_1$] (t5) at (0,-3) {$t_5$};

\node[token] at (p1) {};
\node[token] at (p5) {};

\node[] (name) at (0,-5) {$\netn_1^A = \netn_A \times \netn_B^f$};

\draw[->] (p1) to (t1);
\draw[->] (p1) to (t2);

\draw[<-,bend right = 20] (t3) to (p2);
\draw[<-, bend right = 20] (t4) to (p3);
\draw[->,bend left = 20] (t3) to (p2);
\draw[->, bend left = 20] (t4) to (p3);
\draw[->] (p5) to (t5);

\draw[->] (t1) to (p2);
\draw[->] (t2) to (p3);

\draw[<-,bend right = 20] (t3) to (p4);
\draw[<-, bend right = 20] (t4) to (p4);
\draw[->,bend left = 20] (t3) to (p4);
\draw[->, bend left = 20] (t4) to (p4);
\draw[->] (t5) to (p4);

\end{tikzpicture}}} 
%
   \subfigure{\scalebox{1}{\begin{tikzpicture}[->,>=stealth',shorten >=1pt,auto,node distance=2.5cm,semithick]
\tikzstyle{every state}=[draw]

\node[state] (p1) {};
\node[state] (p2) at (1,0) {};
\node[state] (p4) at (2,0) {};
\node[state] (p5) at (4,1) {};
\node[state] (p7) at (4,0) {};
\node[state] (p8) at (4,-2) {};
\node[state] (p9) at (6,1) {};
\node[state] (p10) at (6,-2) {};

\node[transition,label=above:$o_1$] (t1) at (1,1) {$t_1$};
\node[transition,label=above:$u_3$] (t4) at (5,1) {$t_4$};
\node[transition,label=right:$o_5$] (t5) at (6,0) {$t_5$};
\node[transition,label=below:$o_2$] (t6) at (1,-1) {$t_6$};
\node[transition,label=above:$f$] (t7) at (3,0) {$t_7$};
\node[transition,label=left:$o_3$] (t8) at (4,-1) {$t_8$};
\node[transition,label=below:$u_2$] (t9) at (5,-2) {$t_9$};
\node[transition,label=right:$o_4$] (t10) at (6,-1) {$t_{10}$};

\node[token] at (p1) {};
\node[token] at (p2) {};

\node[] (name) at (3.5,-3) {$\netn_2^D = \netn_C^f \times \netn_D$};

\draw[->] (p1) to (t1);
\draw[->] (p2) to (t1);
\draw[->] (t1) to (p5);
\draw[->] (p5) to (t4);
\draw[->] (t4) to (p9);
\draw[<-, bend right = 20] (p9) to (t5);
\draw[->, bend left = 20] (p9) to (t5);

\draw[->] (p1) to (t6);
\draw[->] (p2) to (t6);
\draw[->] (t6) to (p4);
\draw[->] (t6) to (p8);
\draw[->] (p4) to (t7);
\draw[->] (t7) to (p7);
\draw[<-, bend right = 20] (p7) to (t8);
\draw[->, bend left = 20] (p7) to (t8);
\draw[<-, bend right = 20] (p8) to (t8);
\draw[->, bend left = 20] (p8) to (t8);
\draw[->] (p8) to (t9);
\draw[->] (t9) to (p10);
\draw[<-, bend right = 20] (p10) to (t10);
\draw[->, bend left = 20] (p10) to (t10);

\end{tikzpicture}}}
   \caption{Nets after removing faults in some components}
   \label{fig:pn_reduced}
 \end{figure}

We now consider the net $\netn^i$ composed by component $\netn_{A_i}$ and the fault free version of $\netn_{A_j}$ for $j \not = i$. Figure~\ref{fig:reduced} shows the four components after removing fault $f$ and Figure~\ref{fig:pn_reduced} shows the product nets obtained after these reductions.

\begin{exmp}
  Let us consider the nets from Figure~\ref{fig:pn_reduced}. System $\netn_1^B$ is trivially diagnosable. In the case of $\netn_1^A$, it is easy to see that the observable traces are of the form $\widehat{o}_3$, but all traces containing $o_3$ also contain $f$ and therefore $\netn_1^A$ is also diagnosable. Traces $o_2 u_2 \widehat{o}_4$ and $o_2 f u_2 \widehat{o}_4$ of net $\netn_2^D$ have the same observability, but one contains a fault and the other does not. We can conclude that $\netn_2^D$ is not diagnosable. This result is consistent with the one obtained by the verifier in Example~\ref{ex:veri}.
  \label{ex:two}
\end{exmp}

Clearly the traces of $\netn^i$ are those of $\netn$ such that its projections on every $A_j$ are fault free for $j \not = i$ .

\begin{prop}
  Let $\netn$ be a net, then $\sigma \in \inftraces{\netn^i}$ iff $\sigma \in \inftraces \netn \land \forall j \not = i, \sigma_j: \proj j \sigma = \sigma_j \imp f \not \in \sigma_j$.
  \label{prop:traces}
\end{prop}

The following result states necessary conditions for the diagnosability of $\netn$, i.e. the non diagnosability of $\netn^i$ for some $i$ implies the non diagnosability of $\netn$.

\begin{thm}
  Consider the net $\netn$, then $$\diag \netn \imp \bigwedge\limits_{i=1}^n\diag{\netn^i}$$
  \label{the:one}
\end{thm}

\begin{pf}
  Lets assume that $\neg \diag{\netn^i}$ for some $i$, then there exist $\sigma, \alpha \in \inftraces{\netn^i}$ and $f$ such that $\obs \sigma = \obs \alpha$ with $f \in \sigma$, but $f \not \in \alpha$. We know from Proposition~\ref{prop:traces} that every trace in $\netn^i$ is a trace in $\netn$, so we have found two traces of $\netn$ with the same observability, one containing a fault and the other one not. Therefore $\netn$ is non-diagnosable.
\end{pf}

\begin{exmp}
  We see in Example~\ref{ex:two} that $\netn_2^D$ is non diagnosable. Using Theorem~\ref{the:one} we can conclude that $\netn_2$ is non diagnosable, which is consistent with the diagnosability analysis made in Example~\ref{ex:one}.
\end{exmp}

As explained above, the idea is to build a diagnosable component and to test that its interaction with the others fault free component is also diagnosable. We can then decide the diagnosability of $\netn = \netn_{A_1} \times \dots \times \netn_{A_n}$ in terms of the diagnosability of $A_i$ and $\netn^i$.

\begin{thm}
  Let $\netn = \netn_{A_1} \times \dots \times \netn_{A_n}$, then $$\bigwedge\limits_{i=1}^n (\diag{A_i} \land \diag{\netn^i}) \imp \diag \netn$$
\end{thm}

\begin{pf}
  Let assume that we have a fault $f \in \Sigma_F$ and $\sigma, \alpha \in \inftraces \netn$ with $f \in \sigma$ and $\obs \sigma = \obs \alpha$, we need to prove that $f \in \alpha$. Consider the following cases: 
  \begin{enumerate}
    \item if $\sigma, \alpha \in \inftraces{\netn^i}$ we can prove by $\netn^i$'s diagnosability that $f \in \alpha$ and then $\netn$ is diagnosable,
    \item if $\alpha \not \in \inftraces{\netn^i}$, using the hypothesis that $\alpha \in \inftraces \netn$, we can apply Proposition~\ref{prop:traces} and obtain that $\exists \alpha_i: \proj i \alpha = \alpha_i \land f \in \alpha_i$. By Proposition~\ref{prop:fault_proj} we know that every fault belonging to a projection also belongs to the trace in the net, then $f \in \alpha$ and $\netn$ is diagnosable,
    \item if $\alpha \in \inftraces{\netn^i}$ and $\sigma \not \in \inftraces{\netn^i}$ we know by Proposition~\ref{prop:traces} that $\forall \alpha_i: \proj i \alpha = \alpha_i$ and $f \not \in \alpha_i$ and also that $\exists \sigma_i : \proj i \sigma = \sigma_i$ with $f \in \sigma_i$. As $\obs \sigma = \obs \alpha$ we have that $\obs{\sigma_i} = \obs{\alpha_i}$ by Proposition~\ref{prop:obs}. Finally, as $A_i$ is diagnosable and $f \in \sigma_i$, the fault should belong to $\alpha_i$, leading to a contradiction. We can conclude that $\netn$ is diagnosable.
  \end{enumerate}
\end{pf}

\section{Conclusions} \label{sec:CFW}

We have presented a framework for the distributed diagnosability analysis of concurrent systems. We remove the assumption that a kind of fault can only occur in a single component (which is usually made in faulty distributed systems) and allow to analyze more general systems. The method presented in this paper is a continuation of ~\cite{par_diag}, which to the best of our knowledge, is the first method that allows the diagnosability analysis to be done in a parallelized manner. Thus, a component can do the diagnosability analysis independently of other components, even when the other components are not yet ready. Furthermore, we employ LTL-X model checking based on Petri net unfolding to test diagnosability, which has been proven to be very efficient.

We plan to try to reduce the system in order to obtain minimal components from which we can infer the diagnosability of the original global system. In addition, we intend to relax the assumption that the communicating (synchronizing) events are observable. Moreover, we aim to apply our framework to other diagnosability related properties such as predictability.

\bibliography{ifacconf}
\bibliographystyle{named}

\end{document}